\documentclass[final]{iacrtrans}
\pdfoutput=1
\usepackage[utf8]{inputenc}
\usepackage[english]{babel}

\usepackage{url}
\usepackage{hyperref} % Needed before loading 'cleveref'

\usepackage{amsmath,amssymb,amsthm,wasysym,mathtools,esvect}

\usepackage[babel=true]{csquotes}
\usepackage{cleveref}

\usepackage{amsmath,amssymb,amsthm,wasysym,mathtools,esvect,dsfont}

\usepackage{graphicx}
\usepackage{color}
\makeatletter
\@ifpackageloaded{xcolor}{}{
  \usepackage[dvipsnames]{xcolor}
}

\definecolor{persianrose}{rgb}{1.0, 0.16, 0.64}
\definecolor{goldenyellow}{rgb}{1.0, 0.87, 0.0}

\makeatother

\usepackage{float}

\usepackage{tikz}
\usetikzlibrary{automata}
\usetikzlibrary{arrows}
\usetikzlibrary{shadows}
\usetikzlibrary{decorations.text}
\usetikzlibrary{circuits.logic.US} 
\usetikzlibrary{matrix}

\usepackage{pgfplots}

\usepackage{subcaption}

\usepackage{enumitem}

\usepackage{environ}

\usepackage{tabularx}
\usepackage{makecell}
\usepackage{multirow}
\usepackage{colortbl}
\usepackage{collcell}
\usepackage{threeparttable}
\usepackage{adjustbox}
\usepackage{booktabs}

\usepackage{listings}
\let\origthelstnumber\thelstnumber
\makeatletter
\newcommand*\Suppressnumber{%
  \lst@AddToHook{OnNewLine}{%
    \let\thelstnumber\relax%
     \advance\c@lstnumber-\@ne\relax%
    }
}

\newcommand*\Reactivatenumber[1]{%
  \setcounter{lstnumber}{\numexpr#1-1\relax}
  \lst@AddToHook{OnNewLine}{%
   \let\thelstnumber\origthelstnumber%
   \refstepcounter{lstnumber}
  }
}

\makeatother
\lstdefinestyle{vhdl_style}
{
    language=VHDL,
    float=!htb,
    basicstyle=\ttfamily\footnotesize,
    identifierstyle=\bfseries\color{black},
    keywordstyle=\bfseries\color{persianrose},
    stringstyle=\bfseries\color{yellow},
    commentstyle=\bfseries\color{gray},
    columns=flexible,
    frame=single,
    showspaces=false,
    showstringspaces=false,
    numberstyle=\tiny,
    stepnumber=1,
    breaklines=true,
    xrightmargin=-\fboxsep,
    backgroundcolor=\color{white},
    captionpos=t,
    mathescape,
    escapechar=\%
}

\usepackage{algorithm}
\usepackage[noend]{algpseudocode}
\algrenewcommand\algorithmicrequire{\textbf{\ \ Input:}}
\algrenewcommand\algorithmicensure{\textbf{Output:}}

\usepackage{microtype}
\usepackage{xspace}
\usepackage{pifont}
\usepackage{comment}

\usepackage{ulem}

\usepackage[nolist]{acronym}
\begin{acronym}

\acro{3DES}{Triple-DES}

\acro{AES}{Advanced Encryption Standard}
\acro{ALU}{Arithmetic Logic Unit}
\acro{API}{Application Programming Interface}
\acro{ARX}{Addition Rotation XOR}
\acro{ASIC}{Application Specific Integrated Circuit}
\acro{ASIP}{Application Specific Instruction-Set Processor}
\acro{AS}{Active Serial}

\acro{BDD}{Binary Decision Diagram}
\acro{BGL}{Boost Graph Library}
\acro{BNF}{Backus-Naur Form}
\acro{BRAM}{Block-Ram}

\acro{CBC}{Cipher Block Chaining}
\acro{CFB}{Cipher Feedback Mode}
\acro{CFG}{Control Flow Graph}
\acro{CLB}{Configurable Logic Block}
\acro{CLI}{Command Line Interface}
\acro{COFF}{Common Object File Format}
\acro{CPA}{Correlation Power Analysis}
\acro{CPU}{Central Processing Unit}
\acro{CRC}{Cyclic Redundancy Check}
\acro{CTR}{Counter}

\acro{DC}{Direct Current}
\acro{DES}{Data Encryption Standard}
\acro{DFA}{Differential Frequency Analysis}
\acro{DFT}{Discrete Fourier Transform}
\acro{DLL}{Dynamic Link Library}
\acro{DMA}{Direct Memory Access}
\acro{DNF}{Disjunctive Normal Form}
\acro{DPA}{Differential Power Analysis}
\acro{DSO}{Digital Storage Oscilloscope}
\acro{DSP}{Digital Signal Processing}
\acro{DUT}{Design Under Test}

\acro{ECB}{Electronic Code Book}
\acro{ECC}{Elliptic Curve Cryptography}
\acro{EEPROM}{Electrically Erasable Programmable Read-only Memory}
\acro{EMA}{Electromagnetic Emanation}
\acro{EM}{electro-magnetic}

\acro{FFT}{Fast Fourier Transformation}
\acro{FF}{Flip Flop}
\acro{FI}{Fault Injection}
\acro{FIR}{Finite Impulse Response}
\acro{FPGA}{Field Programmable Gate Array}
\acro{FSM}{Finite State Machine}

\acro{GUI}{Graphical User Interface}

\acro{HDL}{Hardware Description Language}
\acro{HD}{Hamming Distance}
\acro{HF}{High Frequency}
\acro{HSM}{Hardware Security Module}
\acro{HW}{Hamming Weight}

\acro{IC}{Integrated Circuit}
\acro{IO}[I/O]{Input/Output}
\acro{IOB}{Input Output Block}
\acro{IOT}[IoT]{Internet of Things}
\acro{IP}{Intellectual Property}
\acro{ISA}{Instruction Set Architecture}
\acro{IV}{Initialization Vector}

\acro{JTAG}{Joint Test Action Group}

\acro{KAT}{Known Answer Test}

\acro{LFSR}{Linear Feedback Shift Register}
\acro{LSB}{Least Significant Bit}
\acro{LUT}{Look-up table}

\acro{MAC}{Message Authentication Code}
\acro{MIPS}{Microprocessor without Interlocked Pipeline Stages}
\acro{MMIO}{Memory Mapped \acl{IO}}
\acro{MSB}{Most Significant Bit}

\acro{NASA}{National Aeronautics and Space Administration}
\acro{NSA}{National Security Agency}
\acro{NVM}{Non-Volatile Memory}

\acro{OFB}{Output Feedback Mode}
\acro{OISC}{One Instruction Set Computer}
\acro{ORAM}{Oblivious Random Access Memory}
\acro{OS}{Operating System}

\acro{PAR}{Place-and-Route}
\acro{PCB}{Printed Circuit Board}
\acro{PC}{Personal Computer}
\acro{PS}{Passive Serial}
\acro{PUF}{Physical Unclonable Function}

\acro{RISC}{Reduced Instruction Set Computer}
\acro{RNG}{Random Number Generator}
\acro{ROM}{Read-Only Memory}
\acro{ROP}{Return-oriented Programming}
\acro{RTL}{Register Transfer Level}

\acro{SCA}{Side-Channel Analysis}
\acro{SHA}{Secure Hash Algorithm}
\acro{SNR}{Signal-to-Noise Ratio}
\acro{SPA}{Simple Power Analysis}
\acro{SPI}{Serial Peripheral Interface Bus}
\acro{SRAM}{Static Random Access Memory}

\acro{UART}{Universal Asynchronous Receiver Transmitter}
\acro{UHF}{Ultra-High Frequency}

\acro{WISC}{Writeable Instruction Set Computer}

\acro{XDL}{Xilinx Description Language}
\acro{XTS}{XEX-based Tweaked-codebook with ciphertext Stealing}

\end{acronym}

\definecolor{lightcoral}{rgb}{0.94, 0.5, 0.5}
\definecolor{lightskyblue}{rgb}{0.53, 0.81, 0.98}

\newcommand{\CC}[1][]{$\text{C\hspace{-.25ex}}^{_{_{_{++}}}}\ifthenelse{\equal{#1}{}}{}{\text{\hspace{-.625ex}#1}}$\xspace}

\newcommand{\circled}[2][]{
  \tikz[baseline=(char.base)]{
    \node[shape=circle,draw,inner sep=1pt,fill=black]
    (char) {\phantom{\ifblank{#1}{#2}{#1}}};
    \node[text=white] at (char.center) {\makebox[0pt][c]{\textbf#2}};}\xspace}
\robustify{\circled}

\makeatletter
\newcommand{\removeifnextchar}[2]{%
    \begingroup
    \ltx@LocToksA{\endgroup#2}%
    \ltx@ifnextchar@nospace{#1}{%
        \def\next{\the\ltx@LocToksA}%
        \afterassignment\next
        \let\scratch= %
    }{%
        \the\ltx@LocToksA
    }%
}

\newcommand{\etal}{\protect\removeifnextchar{.}{et~al.\@ifnextchar.{}{~}}}

\makeatother

\newcommand{\romannumber}[1]{\uppercase\expandafter{\romannumeral#1}}
\usepackage{todonotes}

\ifdefined\algorithmautorefname

\else
  \newcommand{\algorithmautorefname}{Algorithm}
\fi 

\ifdefined\definitionautorefname

\else 
  \newcommand{\definitionautorefname}{Definition}
\fi

\addto\extrasenglish{  
  \ifdefined\algorithmautorefname
    \renewcommand{\algorithmautorefname}{Algorithm}
  \fi 
  \ifdefined\definitionautorefname
    \renewcommand{\definitionautorefname}{Definition}
  \fi

}

\newcommand{\HAL}{\textnormal{\textsf{HAL}}\xspace}

\newcommand*{\MinNumber}{0.0}
\newcommand*{\MidNumber}{0.76}
\newcommand*{\MaxNumber}{1.0}

\newcommand{\ApplyGradient}[1]{
  \ifdim #1 pt > \MidNumber pt
    \pgfmathsetmacro{\PercentColor}{max(min(100.0*(#1 - \MidNumber)/(\MaxNumber-\MidNumber),100.0),0.00)} %
    \colorbox{green!\PercentColor!yellow}{#1}
  \else
    \pgfmathsetmacro{\PercentColor}{max(min(100.0*(\MidNumber - #1)/(\MidNumber-\MinNumber),100.0),0.00)} %
    \colorbox{red!\PercentColor!yellow}{#1}
  \fi
}

\newcolumntype{R}{>{\collectcell\ApplyGradient}c<{\endcollectcell}}

\def\mathcolor#1#{\@mathcolor{#1}}
\def\@mathcolor#1#2#3{%
    \protect\leavevmode
    \begingroup
    \color#1{#2}#3%
    \endgroup
}

\pgfplotsset{compat=1.14}

\renewcommand{\etal}[0]{\mbox{et~al.}\xspace}

\usepackage{cleveref}
\usepackage{subcaption}
\usetikzlibrary{shapes.geometric}

\title{Stealthy Opaque Predicates in Hardware - Obfuscating Constant Expressions at Negligible Overhead}

\author{Max~Hoffmann \and Christof~Paar}

\institute{Horst G\"ortz Institute for IT-Security, Ruhr-Universit\"at Bochum, Germany, \email{{max.hoffmann, christof.paar}@rub.de}}

\begin{document}

\maketitle

\keywords{Opaque Predicates \and Obfuscation \and Hardware Reverse Engineering \and Hardware Analysis}

%!TEX root = ../main.tex

\begin{abstract}
    Opaque predicates are a well-established fundamental building block for software obfuscation.
    Simplified, an opaque predicate implements an expression that provides constant Boolean output, but appears to have dynamic behavior for static analysis.
    Even though there has been extensive research regarding opaque predicates in software, techniques for opaque predicates in hardware are barely explored.
    In this work, we propose a novel technique to instantiate opaque predicates in hardware, such that they (1) are resource-efficient, and (2) are challenging to reverse engineer even with dynamic analysis capabilities.
    We demonstrate the applicability of opaque predicates in hardware for both, protection of intellectual property and obfuscation of cryptographic hardware Trojans.
    Our results show that we are able to implement stealthy opaque predicates in hardware with minimal overhead in area and no impact on latency.
\end{abstract}

%!TEX root = ../main.tex

\section{Introduction}
\label{op::sec::intro}
The process of designing, implementing and testing hardware cores is both time consuming and costly.
Thus, in today's hardware design process, not every piece of a product is created anew.
Many products consist partially of \ac{IP}-cores, i.e. third party modules.
The market for \ac{IP}-cores is large with USD 2.45 billion in 2013 \cite{eete:2014:clark} and was recently ``expected to be worth USD 6.22 billion by 2023'' \cite{rm:2017:forecast}.
\ac{IP}-cores are sold as either \textit{soft-cores} in the form of HDL code, \textit{firm-cores} as a netlist, or \textit{hard-cores} as a fixed layout for a certain technology.
Preferably they are sold as hard- or firm-cores with only the interfaces and functional specifications known to maintain exclusivity and hinder IP-theft.
Consequently, the \ac{IP}-core market suffers from copyright infringements as code can illegally be copied without relicensing.
Additionally, \ac{IP}-cores have become a major target for reverse engineering in order to understand technologies of competitors and to find hidden backdoors.

Protecting hardware designs has therefore moved into the focus of the design industry.
To counter IP-theft, vendors include watermarks into their products.
A watermark is a vendor- or customer-specific attribute, embedded into the design.
It does not impact functionality but its existence can easily be verified by the vendor.
Consequently, the vendor is able to analyze products on the market for license violations.
However, since such a watermark is often a static part of the design (c.f. \cite{fpl:2006:castillo,schmid:2008:fpt}) or mostly disconnected from the rest of the circuitry (c.f. \cite{el:2003:fan}), it can be found by leveraging these properties and subsequently removed by reverse engineers.
To counter such removal attacks and harden against reverse engineering in general, designers can use obfuscation methods.

Obfuscation in general transforms a product to hide implementation details from reverse engineers while providing functional equivalence.
However, obfuscation often negatively impacts performance.
For example, Chakraborty~\etal applied their technique HARPOON~\cite{chakraborty:2009:tcad} to the ISCAS-89 benchmark suite~\cite{iscas89}, which increased area by $12.44\%$ and power consumption by $12.81\%$ on average.
Since these penalties influence sales, either weak methods are used or no obfuscation is employed at all, with designers arguing that the size of the design makes reversing already infeasible.
Strong obfuscation techniques which can be implemented at negligible costs and with minimal effort could incentivize designers to protect their designs, effectively reducing the amount of information which is plainly available to attackers.

A basic building block in software obfuscation is called \textit{opaque predicate}.
An opaque predicate realizes a function that provides a constant Boolean output known to the designer, but which is difficult to deduce during static analysis.
Opaque predicates can be used, for example, to harden control flow analysis.
However, in hardware obfuscation, such a building block is still missing.
To the best of our knowledge, there has been only a single prior work on dedicated hardware opaque predicates which was shown to be easily detectable and removable (c.f. \Cref{op::sec::background::related-work}).

Many hardware designs incorporate constant control or data signals on HDL level that are worth of protection (e.g. for nominal-actual comparison in sensors or cryptographic key material).
Even though such constant signals may be partially merged into surrounding circuitry by optimization during synthesis, fragmentary information resides which can be leveraged for reverse engineering.
For example, comparison circuitry that evaluates to \textit{true}/\textit{false} for only one input can be realized with fewer gates than a comparator that is \textit{true}/\textit{false} for more than one input, which allows a reverse engineer for easier detection and logical function recovery.
Thus, obfuscating valuable constant signals on HDL level is not only desirable for decreasing the attack surface for reverse engineers, but also for obscuring the functionality of the design post-synthesis.

\paragraph*{Contribution:}
In this work we present stealthy opaque predicates for hardware, a basic building block to obfuscate constant signals.
To this end we discuss criteria for strong opaque predicates, most notably \textit{stealthiness}.
Our technique is independent of the hardware platform (\ac{ASIC} or \ac{FPGA}), requires minimal effort by the designer, and is applicably to the majority of all hardware designs.
We present two implementation strategies, of which one introduces zero additional gates for the instantiation and no additional latency.
Furthermore, we evaluate our technique in two malicious case studies, where we obfuscate backdoors in two cryptographic designs on an \ac{FPGA}, and demonstrate minimal increases in \ac{LUT} and \ac{FF} usage without impacting latency.
Finally, we explain how our opaque predicates can be used to mitigate a known removal attack on a watermarking scheme.

\paragraph*{Outline:}
This work is structured as follows:
\Cref{op::sec::background} provides information on opaque predicates and gives a brief overview on related work in software research.
We also define requirements for our opaque predicates and explain our adversarial model.
In \Cref{op::sec::construction} we explain the general idea behind our opaque predicates, before we present two implementation strategies in \Cref{op::sec::strategies}.
A practical evaluation of the impact on area and latency of our implementation strategies is given in \Cref{op::sec::eval}.
In the end, we show how our opaque predicates can be used to defend a watermarking scheme from a known removal attack in \Cref{op::sec::outlook} before concluding our work in \Cref{sec::conclusion}.

%!TEX root = ../main.tex

\section{Background}
\label{op::sec::background}

\subsection{Static and Dynamic Analysis}
\label{op::sec::background::analysis}
When analyzing a product, two general strategies exist, namely (1) static analysis and (2) dynamic analysis.
While static analysis investigates the product without any interaction, dynamic analysis is performed while the product is operating.
In software reverse engineering, static analysis focuses on the binary or source code, while the code is executed during dynamic analysis.
For hardware reverse engineering, static analysis processes images of a delayered chip, a netlist, or HDL description, while the hardware is in addition also simulated or actually run in dynamic analysis.
The terms are sometimes used interchangeably with offline- and online-analysis.

\subsection{Opaque Predicates}
\label{op::sec::background::op}
Software research has dedicated a lot of effort to opaque predicates.
However, the concept of opaque predicates in hardware has barely been approached yet.
In general, an opaque predicate is a function which has a constant Boolean output, regardless of the input.
This behavior and the constant output is known to the designer, but non-obvious to an observer.
If such a function is found as an expression for a condition in code, static analysis tools cannot be sure which branch is taken.
An example for a simple opaque predicate is the equation ${x(x+1)\stackrel{?}{=} 0 \bmod 2}$ which resolves to \textit{true} for all possible $x$.
Thus, in software, opaque predicates are primarily used for control flow obfuscation.
In hardware, opaque predicates can be used, for example, to obfuscate control signals, constant values, or connections to GND or VCC.

Collberg~\etal published multiple works on software obfuscation using opaque predicates \cite{collberg:2003:gp, collberg:2002:se, collberg:1999:ppl, collberg:1998:ppl}.
Especially their work from 1998 \cite{collberg:1998:ppl} laid the foundation to formalize and evaluate the effectiveness of software opaque predicates.
In the aforementioned paper, the authors present a Java code obfuscator and measure deobfuscation resilience of the generated output.
The main purpose of their obfuscator was to harden control flow analysis by inserting suitable opaque predicates into if statements and loop conditions.
In other words, it tried to make a control flow graph as obscure as possible.
To this end, they presented a variety of techniques to instantiate opaque predicates based on mathematical expressions, as in the example above, pointer equivalence, and concurrency.

Collberg~\etal stated that the quality of obfuscation can be described by four criteria:
the amount of obscurity added to the program (\textit{potency}), the difficulty to break for an automatic deobfuscator (\textit{resilience}), the computational overhead added by the obfuscation (\textit{cost}) and how well the obfuscation blends in with the existing code (\textit{stealthiness}).
We will stick to this terminology when evaluating our own approach.
However, these terms describe evaluation metrics for software obfuscation.
In the following, we briefly explain how these terms translate to hardware designs.

\paragraph{Potency.}
The meaning of \textit{potency} transforms directly to hardware designs.
It indicates how incomprehensible a design becomes, regarding the target(s) of the obfuscation technique.

\paragraph{Cost.}
In software, computational overhead describes longer execution time induced by additional lines of code.
This transfers only partially to hardware, because computations are often done in parallel.
Thus, changes to a design may have no influence on the latency of a design.
In addition, area is typically a more crucial cost factor in hardware designs than code size in software. 
Therefore, cost in hardware measures the overhead in area and latency.

\paragraph{Resilience.}
According to Collberg~\etal, stealthiness is the property of how well an obfuscation blends in with existing code to the ``human eye'', i.e. \textit{manual} inspection, and resilience measures the difficulty of \textit{automated} deobfuscation.
In software analysis, powerful tools exist which automate a majority of the reverse engineering process, hence their output is a suitable indication for resilience.
Dedicated algorithms and intermediate results of the automated tools, for example control flow graphs, give a lot of information for manual inspection to the engineer, making stealthiness assessable.
Therefore, a separation of stealth and resilience makes sense in software.

Regarding hardware, to the best of our knowledge, no publicly available netlist analysis tool with focus on reverse engineering exists.
In 2018, Fyrbiak~\etal presented \HAL, an interactive graph-based netlist analysis tool which enables partial automation of the reversing process and supports the reverse engineers via a dynamic plugin system \cite{fyrbiak:2018:tdsc}.
However, \HAL is not yet available as an open-source project.
Thus, the majority of the hardware reversing process is still performed manually.
Automated deobfuscation can only take place after the manual reversing process revealed the obfuscated parts.
Hence, with resilience in hardware we address the difficulty of automated deobfuscation \textit{after} identification.

Note that once an opaque predicate is detected, it can be further analyzed in order to disclose the constant output.
Since this step is isolated from the remaining hardware, we expect the complexity, i.e. the resilience, to be low in the general case, especially when aided by tools like \HAL.
Therefore, we prioritize stealthiness when evaluating our opaque predicates.

\paragraph{Stealthiness.}
\label{op::sec::background::stealthiness}
As mentioned in the previous paragraph, netlist reverse engineering is mainly performed manually, which in turn makes quantifying stealthiness in hardware even harder.
The \textit{human factor}, i.e. the individual approach of each engineer, is an \textbf{unquantifiable} variable that determines processing speed, reversing skill level, and success rate.
In psychological research, first studies were conducted to examine how human beings engage in reverse engineering.
Lee~\etal had participants deduce the functionality of miniature circuits, for example the functionality of an \texttt{AND}-gate, from a gamified setup with a light bulb and switches \cite{lee:2013:jcp}.
While such studies are an important first step, results are currently too limited to provide quantifiable insights on the human factor in reverse engineering.
Therefore, while it is reasonable to demand high stealthiness from strong opaque predicates, we can only argue the existence of this property for a technique, but we cannot prove it.

Software programmers have a large degree of freedom when writing code.
Even a bunch of additional instructions does not raise suspicion as long as it appears like other code in the project.
Especially instructions related to function calls, math, or Boolean algebra look just like genuine instructions.
In hardware, due to the focus on saving area, extensions to a design are quickly discovered if they lead to a strong deviation from expected size, i.e. there is a correlation with the cost metric.
Since the main functionalities of a design are usually known to the reverse engineer, he also has an idea about the included building blocks.
Additional logical modules in the design can raise suspicion and might lead to identification of obfuscated parts.
Thus, in hardware, stealthiness describes blending not only with genuine circuitry (i.e. code in software), but also with logical modules.

Collberg~\etal also highlighted the context-sensitivity of this metric, as an obfuscation which is regarded stealthy in one project can be highly unstealthy in other projects.
This directly transfers to hardware.

\subsection{Previous Work on Opaque Predicates in Hardware}
\label{op::sec::background::related-work}
In 2011 Meyer-B\"ase~\etal implemented the predicates of Collberg~\etal in Verilog and VHDL, but they did not introduce new constructions \cite{meyer:2011:ica}.
While many publications exist that target opaque predicates in software, we are only aware of a single work regarding dedicated hardware opaque predicates.
It was presented in 2014 by Sergeichik~\etal~\cite{sergeichik:2014:jicms}.
Their approach was to include additional free-running \acp{LFSR} into the design.
An \ac{LFSR} always has a Hamming weight larger than zero, regardless of its current state.
Thus, by feeding all bits of the \ac{LFSR} into an \texttt{OR} gate, the authors generated a constant 1 signal (or 0 signal by using a \texttt{NOR} gate), while the internal \ac{LFSR} state was constantly changing.
However, the technique is attackable because it violates the aforementioned requirement of high stealthiness regarding the logical component.
\acp{LFSR} where all state bits are OR-ed together blend in well with existing circuitry since they are not introducing major overhead in area.
Nevertheless, this logical structure is highly uncommon in hardware designs as it has no other purpose than providing a constant signal.
Exploiting this property, Wallat~\etal presented a practical removal attack on the opaque predicates by Sergeichik~\etal in 2017 \cite{wallat:2017:ivsw}.
They searched for the special \ac{LFSR} structure in a netlist and replaced it with a connection to GND or VCC, in a fully automated fashion.

\subsection{Threat Model}
\label{op::sec::background::model}
We assume the same threat model as prior research on hardware analysis~\cite{alkabani:2007:usenix,chakraborty:2009:tcad,rostami:2014:ieee,bhunia:2014:ieee, book:hardware_obfuscation:chapter1}.
The adversary, i.e. reverse engineer, has full access to a flattened gate-level netlist.
He knows the general functionality and is aware of the input and output specifications, but has no knowledge on internal workings.
His goal is to (1) identify the obfuscated element, (2) analyze, and (3) deobfuscate it.
We emphasize that we especially focus on adversaries which employ static analysis to reverse a design, as this is what opaque predicates are designed to protect against.
However, in \autoref{op::sec::construction::discussion} we argue that our construction even remains stealthy to some extend when faced with dynamic analysis.

Regardless of the employed analysis strategy, an adversary breaks an obfuscation if he succeeds in the third step, i.e. he is able to look through the obscurity added by the obfuscation.
For example, he may want to replace an obfuscated module with a plain equivalent or extract implementation details from an obfuscated design.
To this end, he has to (partially) reverse engineer relevant logical modules in both control and data path.
Note that an ``adversary'' not necessarily has to be malicious.
A designer who tries to detect obfuscated backdoors is also included in our definition.

As demonstrated for numerous settings, it is realistic to assume that the adversary has access to a flattened gate-level netlist:
It can be extracted at chip-level or via layout reverse engineering~\cite{quadir:2016:jetc,vijayakumar:2017:tifs} in the case of \acp{ASIC}, via bitstream-level reverse engineering~\cite{swierczynski:2016:jce,pham:2017:date} in the case of \acp{FPGA}, or is obtained directly from the \ac{IP} provider~\cite{rostami:2014:ieee}.

We emphasize that - given infinite resources - reverse engineering always succeeds, simply because obfuscation must not strip functionality from the design.
This is also formalized in the impossibility result for perfect obfuscation by Barak~\etal \cite{barak:2001:icc}.
However, the cost in both time and effort can quickly grow to extents where reverse engineering is regarded as practically infeasible.
Due to the amount of unknown variables in a design, i.e. the number of gates and the lack of in-depth knowledge on design specifics, we assume that reverse engineering of the full design is out of scope.
In fact, any obfuscation technique would be broken in this case.

\subsection{Requirements}
\label{op::sec::background::requirements}
Our technique is applicable to the majority of all designs.
The only requirement is that it has to include a counter or \ac{FSM}.
We will make use of already existing modules in order to minimize costs.
With more instances of \acp{FSM} or counters in the design stealthiness and resilience of our approach increases.
We regard this requirement as rather weak, since most \ac{IP}-cores contain such building blocks with the exception of, for example, fully combinatorial or pipelined designs.

%!TEX root = ../main.tex

\section{Our Construction}
\label{op::sec::construction}
In this section we explain the high-level idea behind our opaque predicates before we give detailed implementation strategies in \Cref{op::sec::strategies}.

The goal of our construction is to substitute constant signals with \textit{obfuscated constants} which are difficult to detect via static analysis:
An obfuscated constant may change its value arbitrarily throughout the operation of the device, but has a known and stable value at certain periods in time.
In order to achieve high stealthiness, we identify and leverage suitable existing signals or introduce small additional circuitry which we connect to the places where the stable value is needed.

In software, everything which is not affected by the current line of code is static.
Hardware, however, runs in parallel, and signals in the design may change all the time.
The value of a signal is only evaluated if the data or control port it reaches is active, for example the data port of an enabled \ac{FF}.
Therefore, this dynamic switching behavior does not impact functionality if connected ports are deactivated, which is done by control logic.
Consequently, the utilization of obfuscated constants does not change functionality if the stable value is hold as long as succeeding ports are active.
We call this period where the original value has to be available to maintain functionality \textit{processing period}.

The main idea to achieve our goal is to use the state registers of \acp{FSM} or counters as opaque predicates.
Note that these are frequent elements in \ac{IP}-cores, so their inclusion itself does not raise suspicion in contrast to the construction of Sergeichik~\etal \cite{sergeichik:2014:jicms} (c.f. \Cref{op::sec::background::related-work}).
This also explains our requirement that the original circuit has to include such elements.

\acp{FSM} and counters are elements that transition between states based on certain conditions, for example a trigger signal or the previous state.
During the time where no condition is met, the register which encodes the state does not change.
This lays the foundation of our opaque predicates since there are existing elements that have known values at certain periods of operation.
We use such a state register to generate obfuscated constants.

Since \acp{FSM} generate sequences of states analogously to counters which generate sequences of numbers, \acp{FSM} can be regarded as counters with non-numeric counting styles.
For the sake of simplicity we restrict our explanations and examples to \acp{FSM} in the following, but emphasize that presented techniques are similarly applicable to any kind of counters.

Since the sequence of states in an \ac{FSM} cannot be changed without altering device functionality, the designer has to select a suitable \ac{FSM} to ensure a stable signal during the processing period.

\begin{figure}[!htb]
    \centering
    \begin{tikzpicture}
    \tikzstyle{every state}=[fill=white,draw=black,text=black,circular drop shadow]
    %\draw [help lines] (0,0) grid (10,2);

    \node at (-1.25,4) {FSM$_1$};
    \node at (6.5,4) {FSM$_2$};

    \node[initial, state] (startA) at (0,3) {Reset};
    \node[state]  (waitA)     at (0,0)   {Wait};
    \node[state]  (return)    at (-2.5,1.5)   {Send};

    \node[state]                (initA)     at (5,3)    {Init};
    \node[state]                (computeA)  at (6.5,2.5)  {Work1};
    \node[state]                (computeB)  at (8,1.5)  {Work2};
    \node[state]                (computeC)  at (6.5,0.5)  {Work3};
    \node[state, align=center]  (endA)      at (5,0)    {Done};

    \path[->,>=stealth',shorten >=1pt,auto]
    (startA) edge (waitA)
    (waitA) edge [loop below] (waitA)
    (waitA) edge (return)
    (return) edge (startA)

    (initA) edge (computeA)
    (computeA) edge (computeB)
    (computeB) edge [loop right] (computeB)
    (computeB) edge (computeC)
    (computeC) edge (endA)
    ;

    \path[dotted,->,>=stealth',shorten >=1pt,auto]
    (startA) edge node {reset signal} (initA)
    (endA) edge node {done signal} (waitA)
    ;

    \end{tikzpicture}
    \caption{Visualization of two \acp{FSM} which can be used as opaque predicates.}
    \label{op::fig::stable-state}
\end{figure}
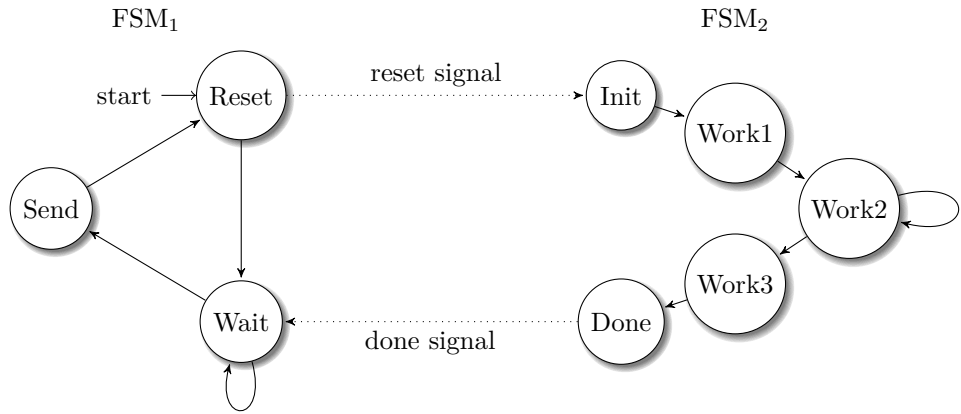

\paragraph*{Example}
\Cref{op::fig::stable-state} shows a small example.
\ac{FSM}$_1$ controls \ac{FSM}$_2$ which in turn controls a computation.
After the computation is done, \ac{FSM}$_1$ triggers transmission of the result.
This process is repeated indefinitely.

If a constant expression is only required by logic which is active in the \textit{Work1}, \textit{Work2}, and \textit{Work3} states of \ac{FSM}$_2$, there are two options to instantiate an opaque predicate:
Over the whole processing period, \ac{FSM}$_1$ remains in the \textit{Wait} state, i.e. the value of the state register is constant and can be used as an opaque predicate.
\ac{FSM}$_2$ passes \textit{Work1}, \textit{Work2}, and \textit{Work3}, so all \acp{FF} of the state register of \ac{FSM}$_2$ which do not change in these states can also be used to generate the constant expression.

\subsection{Discussion}
\label{op::sec::construction::discussion}
We discuss the potential of our opaque predicates according to the criteria originally proposed by Collberg~\etal (c.f. \Cref{op::sec::background::op}).
Furthermore we elaborate on the strength of our opaque predicates when faced with dynamic analysis and inspect the applicability of known techniques against software opaque predicates to our approach.

\paragraph*{Potency.}
With respect to potency, after applying our approach the constant signal is entirely gone when simply ``reading'' the design, i.e. flattened gate-level netlist.
Therefore, regarding the goal of obfuscating constant signals, we introduce maximum obscurity.

\paragraph*{Cost.}
In terms of cost, we do not introduce any additional gates on HDL level when using existing \acp{FSM}.
Depending on the actual instantiation, additional wires are required.
However, as we show in our case studies in \Cref{op::sec::eval}, the synthesis toolchain will add a small number of additional gates, for signal routing and since previous optimization is not possible anymore.
For example, an \texttt{AND} gate where one input is constant can be optimized by the compiler, but as soon as both inputs seem variable, no optimization can be performed and the gate has to be instantiated as is.

\paragraph*{Stealthiness.}
As we argued in \Cref{op::sec::background::stealthiness}, stealthiness cannot be quantified easily.
Therefore, the best we can do is to present arguments why our opaque predicates are stealthy.
Since we utilize existing circuitry, the approach is stealthy in a sense that minimal additional gates and no new \acp{FF} are introduced.
Intuitively, this leads to no unusual structures in contrast to the approach by Sergeichik~\etal (c.f. \Cref{op::sec::background::related-work}).
Our building blocks, \acp{FSM} and counters, are common in \ac{IP}-cores, so their existence is not suspicious either.
Since an \ac{FSM} controls logical modules of a core depending on its current state, signals that reach from a state register into the design are unobtrusive.
Even signals from state registers which are connected to the data inputs of \acp{FF} are common in circuits, as can be seen in our two case studies, which we present in detail in \autoref{op::sec::eval}.
We counted the \ac{FF} instances, where the combinatorial circuit which generated the \ac{FF}'s data input had at least one \ac{FSM} signal as input in relation to all \acp{FF} which receive any signal generated via an \ac{FSM}.
For our (small) PRESENT case study we identified 28.79\% of all \acp{FF}, for our larger RSA case study we identified 37.62\%.
Thus, a reverse engineer cannot use this property to immediately identify obfuscation instances.

Therefore, if an adversary is looking for opaque predicates in general, we are confident that our technique will go unnoticed.
Furthermore, by generating a signal from multiple opaque predicates, i.e. the state registers of multiple \acp{FSM}, the original functionality becomes obstructed even more.
In that case, multiple individual components, which are not related in the first place, contribute to the signal generation, thus requiring a reverse engineer to analyze an even bigger part of the circuit.

However, even if an adversary is looking especially for our opaque predicates we argue that this is only successful with tremendous efforts.
We stress that, while efficient algorithms exist to extract \acp{FSM} from a gate-level netlist as shown by Shi~\etal~\cite{shi:2010:iscas}, this step is not the major difficulty in detecting our opaque predicates.
As mentioned before, \acp{FSM} naturally control other circuitry.
These control signals are outputs of the state register.
The adversary would have to reverse engineer the whole datapath of the logical modules which are controlled by the analyzed \ac{FSM} to detect that a processed signal is actually constant.
To this end, he would also have to deduce the processing period.
We stress that all this additional information, which is required to successfully detect our opaque predicates, tremendously increases the required reversing effort.

Several approaches exist to make sense of a circuits datapath.
In 2012, Shi~\etal presented an automated technique to extract functional modules from a circuit \cite{shi:2012:iscit}, by first stripping all \acp{FSM} from it and subsequently matching subcircuits against abstract models.
At the same time Li~\etal developed a technique based on behavioral patterns to generate a high-level description of unknown circuits \cite{li:2012:host}.
Note that this technique uses dynamic analysis.
One year later, Li~\etal presented another technique to automatically detect word-level \cite{li:2013:host}.
Extending on this technique, Subramanyan~\etal automated extraction of registers and adders \cite{subramanyan:2013:date, subramanyan:2013:tect} and Gasc\'{o}n~\etal improved bit-order reversing using template matching \cite{gascon:2014:fmcad}.
However, all of these techniques only help in recovering modules of the datapath.
Making sense of interconnections and, regarding our construction, identifying processing periods, still has to be performed manually and in conjunction with the control logic. 
Consequently, we regard our opaque predicates as stealthy.

\paragraph*{Resilience.}
As noted in \Cref{op::sec::background::op}, resilience in hardware becomes important once an obfuscation has been identified, i.e. after stealthiness failed.
If an adversary manages to confidently identify an instance of our opaque predicates including the processing period, deobfuscation is straightforward.
By simulating the \ac{FSM} in isolation, i.e. applying a very restricted dynamic analysis, the constant value during the processing period can be disclosed.
Therefore, our approach features low resilience.

\paragraph*{Dynamic Analysis.}
As noted in \Cref{op::sec::background::op} and \Cref{op::sec::background::model}, opaque predicates are designed to defend against static analysis.
However, we want to shortly argue why our construction also offers potential when faced with dynamic analysis.
Since dynamic analysis allows for inspection during design operation via simulation or execution, an adversary would be able to detect signals which do not switch in certain periods of time.
However, this is also (partially) normal behavior of control signals.
Since the \acp{FSM} which are used as opaque predicates also control other parts of the circuit, their signals naturally remain constant for certain periods of time, for example to enable parts of a circuit or to act as state-dependent bitmasks.
Hence, while identifying gates which receive a constant input over a certain duration is possible with dynamic analysis, they do not necessarily have to be the target of an opaque predicate.
Therefore, dynamic analysis is only capable of narrowing the parts of the datapath which have to be analyzed to those including the identified \acp{FF}, but does not instantly break our construction.

\paragraph*{Techniques Against Software Opaque Predicates.}
Multiple approaches exist to attack software opaque predicates.
For example, Dalla~\etal presented an attack based on dynamic analysis to identify software opaque predicates of the form $f(x) = 0 \bmod y$ \cite{dalla2006opaque}.
They observe conditional jumps over a number of executions and propose a technique based on abstract interpretation to detect opaque predicates in these conditions.
Ming~\etal presented LOOP, a \textit{L}ogic-\textit{O}riented \textit{O}paque \textit{P}redicate detection tool using static analysis after a symbolic execution \cite{ming2015loop}.
To this end, they examine all conditions on the execution path, both in isolation as well as in conjunction to detect correlated predicates.

In general, approaches that target software opaque predicates all share a common structure.
First, they identify branch conditions and then further evaluate/compare satisfiability of these conditions in order to identify opaque predicates.
This is a well suited approach in software since opaque predicates are used for control flow obfuscation and do not impact other computations.

We stress that this general approach cannot be plainly transferred to hardware circuits.
In software, control flow is observable from conditional branches, but data flow is obscure by the nature of pointers, while in hardware data flow is observable from wires and combinatorial logic between \acp{FF}, but control flow is obscure because of state machines.
Note that understanding the data flow is distinct from understanding the meaning of the datapath in hardware, it only describes \textbf{where} information goes, not what it means.
Our construction, reuses existing signals which are also used at multiple other places in the circuit, i.e. there is no isolated use for obfuscation.
Therefore, from our point of view, techniques against software opaque predicates cannot be applied to our opaque predicates.

%!TEX root = ../main.tex

\section{Implementation Strategies}
\label{op::sec::strategies}
In this section we present two implementation strategies.
The first one explains in detail how an existing state machine can be used as an opaque predicate.
The second strategy can be used if no existing state machine or counter is suitable for the first method.

\subsection{Implementation in Existing Circuitry}
\label{op::sec::strategies::existing}
In order to use an existing \ac{FSM} as an opaque predicate the designer has to analyze which of the available state registers are suitable.
He first has to identify the processing period $T$, consisting of starting event and duration of the constant to obfuscate.
He then has to verify that the \ac{FSM} passes the same set of states $S$ during each iteration of $T$.
If $S$ changes for several occurrences of $T$, the respective \ac{FSM} cannot be used as an opaque predicate.
This can be the case, for example, if the \ac{FSM} is independent of the circuitry which processes the constant.
If the set of passed states is stable for each iteration of $T$, the state machine is a valid candidate.

After selecting a valid candidate the designer has to configure the state encoding of the \ac{FSM}.
The encodings of all states not in $S$ can be chosen arbitrarily.
For all states in $S$, encodings have to be selected where at least one common bit remains constant.
Consequently, the outputs of the corresponding \acp{FF} remain constant during $T$ and can be used to generate the signals of an obfuscated constant.

Using this implementation strategy no additional \acp{FF} and no combinatorial gates are added to the design.
However, as becomes evident in the case studies in \Cref{op::sec::eval} additional gates may be inferred due to optimization issues.

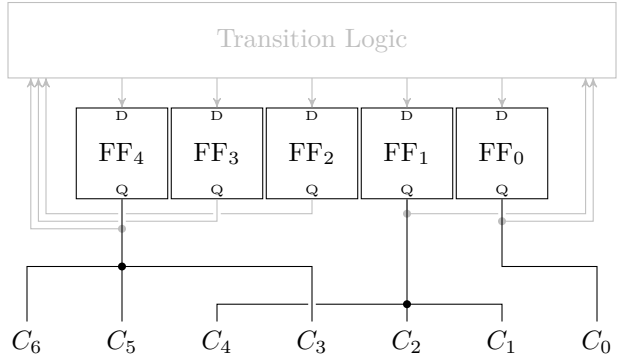
\begin{figure}[!htb]
    \centering
    \begin{tikzpicture}[square/.style={regular polygon,regular polygon sides=4}, gray/.style={draw=lightgray,text=lightgray}]
    \node at (0,2) [square,draw] (FF4) {$\ac{FF}_4$};
    \node at (0,1.5) {\tiny Q};
    \node at (0,2.5) {\tiny D};
    \node at (1.25,2) [square,draw] (FF3) {$\ac{FF}_3$};
    \node at (1.25,1.5) {\tiny Q};
    \node at (1.25,2.5) {\tiny D};
    \node at (2.5,2) [square,draw] (FF2) {$\ac{FF}_2$};
    \node at (2.5,1.5) {\tiny Q};
    \node at (2.5,2.5) {\tiny D};
    \node at (3.75,2) [square,draw] (FF1) {$\ac{FF}_1$};
    \node at (3.75,1.5) {\tiny Q};
    \node at (3.75,2.5) {\tiny D};
    \node at (5,2) [square,draw] (FF0) {$\ac{FF}_0$};
    \node at (5,1.5) {\tiny Q};
    \node at (5,2.5) {\tiny D};

    \node at (-1.25,-0.5)    (C6) {$C_6$};
    \node at (0,-0.5)    (C5) {$C_5$};
    \node at (1.25,-0.5) (C4) {$C_4$};
    \node at (2.5,-0.5)  (C3) {$C_3$};
    \node at (3.75,-0.5) (C2) {$C_2$};
    \node at (5,-0.50)    (C1) {$C_1$};
    \node at (6.25,-0.5) (C0) {$C_0$};

    \draw[<-,>=stealth',gray] (FF0) -- (5,3);
    \draw[<-,>=stealth',gray] (FF1) -- (3.75,3);
    \draw[<-,>=stealth',gray] (FF2) -- (2.5,3);
    \draw[<-,>=stealth',gray] (FF3) -- (1.25,3);
    \draw[<-,>=stealth',gray] (FF4) -- (0,3);

    \draw[->,>=stealth',gray] (5,1.1) -| (6.2,3);
    \fill[fill=lightgray] (5,1.1) circle [radius=1.5pt];

    \draw[gray] (FF1) |- (4.95,1.2);
    \draw[->,>=stealth',gray] (5.05,1.2) -| (6.1,3);
    \fill[fill=lightgray] (3.75,1.2) circle [radius=1.5pt];

    \draw[gray] (FF2) |- (1.3,1.2);
    \draw[gray] (1.2,1.2) -- (0.05,1.2);
    \draw[->,>=stealth',gray] (-0.05,1.2) -| (-1,3);

    \draw[gray] (FF3) |- (0.05,1.1);
    \draw[->,>=stealth',gray] (-0.05,1.1) -| (-1.1,3);

    \draw[->,>=stealth',gray] (0,1) -| (-1.2,3);
    \fill[fill=lightgray] (0,1) circle [radius=1.5pt];

    \node at (2.5,3.5) [rectangle,gray,minimum width=8cm,minimum height=1cm] (FB) {Transition Logic};

    \draw (FF4) -- (C5);
    \fill ($(FF4) - (0, 1.5)$) circle [radius=1.5pt];
    \draw ($(FF4) - (0, 1.5)$) -| (C3);
    \draw ($(FF4) - (0, 1.5)$) -| (C6);

    \draw (FF0) -- ($(FF0) - (0, 1.5)$) -| (C0);

    \draw (FF1) -- (C2);
    \fill (3.75, 0) circle [radius=1.5pt];
    \draw (2.55,0) -| (C1);
    \draw (2.45,0) -| (C4);
    \end{tikzpicture}
    \caption{Exemplary constant generation from a state register.}
    \label{op::fig::implementation::existing::example}
\end{figure}

\paragraph*{Example.}
An \ac{FSM} with a 5-bit state register $(\ac{FF}_4, \ldots, \ac{FF}_0)$ is used as an opaque predicate to generate the 7-bit constant $(C_6, \ldots, C_0) = \texttt{1101000}_2$.
The designer identified that during period $T$ where the constant is required the \ac{FSM} consistently passes three states that form the set $S$.
He encodes them as follows: $S = (\texttt{10100}_2, \texttt{11000}_2, \texttt{11100}_2)$.
Therefore, he can use $\ac{FF}_4$, $\ac{FF}_1$, and $\ac{FF}_0$ which constantly output 1, 0 and 0 during $T$ to generate the obfuscated constant.
In the remaining states of the \ac{FSM} those bits are not fixed and therefore the output appears to have dynamic behavior.
A possible instantiation would be to connect these \acp{FF} as shown in \Cref{op::fig::implementation::existing::example}.

\subsection{Implementation with Additional Circuitry}
\label{op::sec::strategies::additional}
If the designer is in a situation where none of the existing \acp{FSM} or counters offer a stable set $S$ during the processing period, he can still make use of our general idea.
This situation can for example occur if the design consists of independent \acp{FSM} where the state sequence during the processing period changes depending on external events.
In this section we describe an implementation strategy which adds a small number of gates and \acp{FF} in order to create a module that looks like a genuine \ac{FSM}.
However, the new \ac{FSM} will reach a known stable state after a number of clock cycles, thus providing constant values after the initial delay.
The designer only has to add the \ac{FSM} to the design and connect the reset signal to be triggered at a point which is far enough ahead of the processing period.
Note that generating an \ac{FSM} which cycles a known sequence of states during the processing period is highly difficult in this scenario. 
In the best case, we are only able to give a lower bound for the time difference between reset and processing period.
By making sure that the additional \ac{FSM} is in its stable state simply before the processing period is reached, we do not have to care for varying amounts of time after \ac{FSM} reset.

The difficulty in generating this opaque predicate is that we have to find interconnections for the state register's \acp{FF} which result in a stabilization from a known starting value, i.e. the reset value, to a known stable value after a known number of clock cycles.
We present two algorithmic approaches which lead to three algorithms \texttt{QM}, \texttt{QMX}, and \texttt{RND} that automatically generate such interconnections.
They require the register size $n$ and start/reset value $s$ as input as well as a stabilization delay $t$.
The algorithms output the stable state $z$ and the corresponding function $f : \{0,1\}^n \rightarrow \{0,1\}^n$ which represents the interconnections.
Note that all algorithms can be adjusted to regard the input $t$ as a lower bound $t_{min}$ and internally select a random $t \geq t_{min}$.
In that case, the actual stabilization delay $t$ becomes an additional output.

The transition function of the \ac{FSM} during each clock cycle is then described by the function ${F : \{0,1\}^n \times \{0,1\} \rightarrow \{0,1\}^n}$ where
$$F(x,RST) = \begin{cases} s & RST=1 \\ f(x) & else \end{cases}$$
i.e. one clock cycle in the hardware implementation updates the register from state $x$ to state $f(x)$ or resets the register if required.
In the following we first present the three algorithms before comparing their outputs.

\paragraph*{Approach 1: Quine-McCluskey Supported Generation.}
Our first approach chooses a random sequence of states and utilizes Boolean algebra to generate an according transition function $f$.
The chosen sequence $X = (x_i)_{0 \leq i \leq t+1}$ is of length $t+2$, where $x_0$ is set to $s$ and $x_1, \ldots, x_{t-1}$ are chosen randomly such that no state appears twice.
The two final states in the sequence have to be equal, i.e. $x_{t} = x_{t+1}$.
When $f$ is generated based on $X$, it outputs said sequence, thus $f(x_i) = x_{i+1}$ holds and subsequently $f(x_{t}) = x_{t+1} = x_{t}$, i.e. the function stabilizes after $t$ clock cycles in state $z = x_{t}$.

We use the Quine-McCluskey algorithm to generate minimized Boolean functions for each bit separately \cite{quine:1952:tamm}.
The Quine-McCluskey algorithm basically transforms a truth table into the corresponding Boolean function.
The truth table can also contain so called ``don't care states'' where the output is irrelevant for the respective input values.
The algorithm requires a set of values which shall result in a 1 at the output and the ``don't care states''.
It then returns a Boolean function in \ac{DNF} which outputs a 1 for all specified inputs, either 1 or 0 for all ``don't care'' values and a 0 for all other inputs.
In our case, all states which are not in the sequence are ``don't care states''.

The original Quine-McCluskey algorithm outputs a function in \ac{DNF}, i.e. it only uses \texttt{AND}, \texttt{OR}, and \texttt{NOT} gates.
However, modifications exist which also make use of \texttt{XOR} gates.
In the following, we refer to the above algorithm that uses the original Quine-McCluskey as \texttt{QM} and to the one using the XOR-enhanced variant as \texttt{QMX}.

\begin{figure}[!htb]
    \centering
    \usetikzlibrary{arrows, shapes.gates.logic.US, calc}
    \begin{tikzpicture}
    \node at (-1,3) (FF0) {FF$_0$};
    \node[text = gray] at (-1,2.25) (FF1) {FF$_1$};
    \node at (-1,1.5) (FF2) {FF$_2$};
    \node at (-1,0.75) (FF3) {FF$_3$};
    \node[text = gray] at (-1,0) (FF4) {FF$_4$};

    \node[not gate US, draw] at (1,3) (notx) {};
    \node[or gate US, draw, rotate=0, logic gate inputs=nnn] at (4,1.5) (gate) {};

    \draw[dotted] (-0.25,3.75) -- (-0.25,-0.25);
    \node at (1,3.6) {Inverter Stage};
    \draw[dotted] (2.25,3.75) -- (2.25,-0.25);

    \draw (FF0) -- (notx.input);
    \draw (notx.output) -- (3,3) |- (gate.input 1);

    \draw (FF2) -- (3,1.5) |- (gate.input 2);
    \draw (FF3) -- (3,0.75) |- (gate.input 3);
    \draw (gate.output) -- (5,1.5);

    \node[anchor=west] at (5, 1.5) {output};

    \end{tikzpicture}
    \caption{Exemplary interconnection as generated by the \texttt{RND} algorithm.}
    \label{op::fig::implementation::additional::example}
\end{figure}
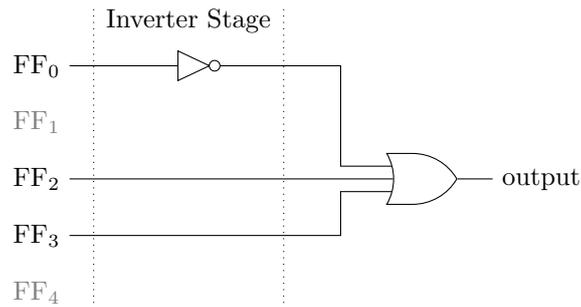

\paragraph*{Approach 2: Randomized Generation.}
\label{op::sec::algorithm::random}
In contrast to our first approach, the second one generates a random transition function $f$ for an $n$-bit state register and then simulates the sequence of generated states in order to determine suitability.
The algorithm, further denoted as \texttt{RND}, also processes each bit individually when generating $f$.
For each \ac{FF} in the register, it first selects a random subset $I$ of all other \acp{FF} of said register, where $|I| \geq 2$ has to be ensured.
The output signals of all \acp{FF} in $I$ are put through an inverter stage where a randomly selected subset is negated.
The output signals of the inverter stage are then passed as inputs to an \texttt{OR}, \texttt{AND}, or \texttt{XOR} gate, which is again randomly selected.
The output of that gate is the output bit of $f$, which is used to update the currently processed \ac{FF}.
An exemplary interconnection for a single \ac{FF} of a 5-bit state register $(\ac{FF}_4, \ldots,  \ac{FF}_0)$ as generated by this algorithm is shown in \Cref{op::fig::implementation::additional::example}.
In this case $I = \{\ac{FF}_0, \ac{FF}_2, \ac{FF}_3\}$ was chosen, the output of $\ac{FF}_0$ was selected to be the input of a \texttt{NOT} gate in the inverter stage, and an \texttt{OR} gate was instantiated to finally combine the signals.

After generating $f$ as described above, simulation starts at state $s$ and simulates until a stable state $z$ is found.
If the number of clock cycles until $f$ stabilizes in state $z$ is $\geq t$, the algorithm outputs $f$ and $z$.
Otherwise a new function $f$ is chosen and the process is repeated.

\paragraph*{Discussion.}
\label{op::sec::strategies::additional::discussion}
Our algorithms tackle the same task from different directions.
While \texttt{QM} and \texttt{QMX} take a bottom-up approach of selecting a state sequence and generating the corresponding transition function, \texttt{RND} takes the top-down approach of selecting a transition function and verifying its suitability.

Since opaque predicates are generated during the design process, we regard runtime as an insignificant property of our algorithms.
For the sake of completeness we shortly elaborate on the respective expected execution times.
For a truth table with $k$ entries Quine-McCluskey terminates after $\mathcal{O}(3^k \ln(k))$ steps (worst-case), therefore \texttt{QM} and \texttt{QMX} terminate in a similar amount of steps.
Since \texttt{RND} solves the same problem, it terminates as well, but accurate runtime estimations are difficult because of its random nature.
According to our experiments, if $t$ gets closer to $2^n$, the maximum number of states an $n$-bit register can encode, the random search can become very long.

\begin{table}
    \centering
    \begin{tabular}{cc|ccc|cccc|cccc}
    	                              &    $n$    & \multicolumn{3}{c|}{3} &  \multicolumn{4}{c|}{4}  &   \multicolumn{4}{c}{5}   \\
    	                              & $t$ &  2  &  3  &     5      &  2  &  3   &  5   &  10  &  2   &  3   &  5   &  10  \\ \hline
    	\multirow{4}{*}{\texttt{QM}}  &   \#not   & 2.6 & 2.9 &    3.0     & 3.5 & 3.8  & 4.0  & 4.0  & 4.3  & 4.7  & 5.0  & 5.0  \\
    	                              &   \#and   & 3.8 & 4.3 &    6.1     & 6.7 & 8.3  & 12.8 & 24.6 & 10.0 & 13.6 & 22.0 & 45.7 \\
    	                              &   \#or    & 5.1 & 5.4 &    5.3     & 9.5 & 11.3 & 13.7 & 16.0 & 14.6 & 19.0 & 25.2 & 34.5 \\
    	                              &   \#xor   & 0.0 & 0.0 &    0.0     & 0.0 & 0.0  & 0.0  & 0.0  & 0.0  & 0.0  & 0.0  & 0.0  \\ \hline
    	\multirow{4}{*}{\texttt{QMX}} &   \#not   & 1.9 & 2.4 &    2.9     & 2.3 & 3.0  & 3.9  & 5.4  & 2.3  & 3.3  & 4.5  & 6.9  \\
    	                              &   \#and   & 0.7 & 1.4 &    3.0     & 0.8 & 1.8  & 4.8  & 13.5 & 0.9  & 2.2  & 6.6  & 20.9 \\
    	                              &   \#or    & 5.1 & 5.6 &    5.9     & 9.1 & 11.3 & 15.0 & 21.6 & 13.4 & 18.3 & 26.5 & 43.7 \\
    	                              &   \#xor   & 1.8 & 1.9 &    2.3     & 3.2 & 4.0  & 5.8  & 9.0  & 4.9  & 6.8  & 10.8 & 19.0 \\ \hline
    	\multirow{4}{*}{\texttt{RND}} &   \#not   & 2.4 & 2.5 &     -      & 3.3 & 3.3  & 3.2  & 3.0  & 4.3  & 4.2  & 4.2  & 4.1  \\
    	                              &   \#and   & 1.3 & 1.2 &     -      & 2.3 & 2.2  & 1.8  & 0.8  & 3.7  & 3.5  & 3.0  & 1.8  \\
    	                              &   \#or    & 1.7 & 1.8 &     -      & 3.6 & 3.6  & 3.8  & 4.1  & 5.8  & 6.0  & 6.2  & 6.8  \\
    	                              &   \#xor   & 0.5 & 0.5 &     -      & 1.3 & 1.5  & 1.9  & 3.3  & 2.3  & 2.6  & 3.3  & 4.9
    \end{tabular}
    \caption{Comparison of additional gates required for the interconnections of state register \acp{FF}.}
    \label{op::alg::output}
\end{table}

All three algorithms have to add additional circuitry, namely $n$ \acp{FF} and the gates required to realize $f$.
\Cref{op::alg::output} shows the amount of gates each algorithm requires to generate $f$, averaged over 1000 executions for each parameter set.
Since $f$ is generated independently for each \ac{FF}, there may be common terms in the Boolean functions.
While a compiler could optimize this during synthesis, we employed a postprocessing step directly on $f$ to extract common terms which have to only be computed once in order to get a fair comparison in \Cref{op::alg::output}.
We explain the parameter choices later when we extend on stealthiness, as both topics are related.

For the \texttt{RND} algorithm we did not compute values for $n=3, t=5$.
With $n=3$ only 8 states are possible and randomly finding $f$ that stabilize after $\geq 5$ clock cycles took minutes to hours, while results of the first hundred iterations indicated no different behavior compared to the other inputs.
It is easy to see that \texttt{RND} introduces fewer additional gates than the output of the Quine-McCluskey supported algorithms.
Furthermore, the amount of gates required for \texttt{RND} is mainly influenced by the choice of $n$ while for \texttt{QM} and \texttt{QMX} $n$ and $t$ both notably impact the gate count.
The advantage of \texttt{RND} becomes more significant for bigger values of $n$ and $t$.

All algorithms can be tweaked to use different strategies for handling the input $t$, as for example it could be interpreted as a lower bound $t_{min}$ for additional randomness.
However, \texttt{RND} offers additional degrees of freedom for the actual implementation:
When generating a random $f$, the strategy of creating connections can be adapted, e.g. more or less inputs can be considered for each bit or multiple gates can be inserted.
The output of the Quine-McCluskey method is always in \ac{DNF}, a structure where numerous \texttt{AND} gates are input to \texttt{OR} gates.
An \ac{FSM} where the state transition logic is in \ac{DNF} structure could offer an anchor point for detection to a reverse engineer.

This leads to our conclusion that the randomized generation returns much smaller and thus more natural, i.e. more stealthy, interconnections and can be customized to a larger extend than the Quine-McCluskey supported generation.
Therefore, we regard the randomized approach as the algorithm of choice.

We stress that regardless of the employed algorithm, there is a stabilization delay.
Depending on the point in time where the reset signal is triggered, the designer might have to infer an artificial delay to ensure that the \ac{FSM} stabilized before the processing period.

\paragraph*{Stealthiness and Parameter Choice.}
Since we add new circuitry to the existing design with this approach, stealthiness is lower than for the first variant.
Still the inclusion of an additional \ac{FSM} in a design with multiple benign \acp{FSM} is not suspicious in the first place.
In order to integrate the \ac{FSM} into the whole design and not make it appear as an outsider, it should be reset by other state machines.
That way it gets integrated into the control hierarchy and blends in better than an autonomous module.
To further improve similarities with existing \acp{FSM}, parameters $n$ and $t$ should be selected to output \acp{FSM} which do not vastly differ from other \acp{FSM} in the design in terms of size.
For example could an \ac{FSM} with a 15-bit state register be suspicious in a design where each module only has a low amount of logical states.
Because of this \Cref{op::alg::output} was created with $n \in \{3,4,5\}$ as these were common \ac{FSM} register sizes we experienced.

Stealthiness is only barely affected by $t$.
Even for small values, the generated interconnections are not suspicious during static analysis.
Only during simulation, i.e. dynamic analysis, larger values mean that the stabilizing behavior will occur later.
Therefore, the designer can select a value of $t$ which allows for easy integration.
To map this property we selected arbitrary values in \Cref{op::alg::output} to only analyze the impact on additional gates.

Static analysis, presented with a gate-level netlist, cannot determine from the wiring that the \ac{FSM} produces a stable signal after a number of clock cycles from reset.
Only via simulation this behavior becomes evident.
However, even if this property is uncovered, the opaque predicate is not immediately identified as such.
\acp{FSM} often control a sequence of operations before idling in a \textit{Done} state, i.e. even genuine \acp{FSM} may stabilize in a certain state.
In order to actually identify our opaque predicate \acp{FSM}, still further reverse engineering of the influenced datapath is required, immensely increasing the attacker's efforts.

The technique with additional circuitry introduces some potential weaknesses.
The new \ac{FSM} is mostly self-contained, i.e. only the reset is performed by external control signals.
While such \acp{FSM} \textit{can} be a normal part of some circuits, they might be a suspicious module in others, resulting in a comparable weakness to the approach of Sergeichik~\etal \cite{sergeichik:2014:jicms} (c.f. \autoref{op::sec::background::related-work}).
However, without further inspection our modules still look just like genuine \acp{FSM} in the general case and require an adversary to invest additional effort to detect.

We therefore conclude that our opaque predicates with additional circuitry still remain stealthy, although they are weaker than our opaque predicates in existing circuitry.

%!TEX root = ../main.tex

\subsection{Practical Evaluation}
\label{op::sec::eval}
In order to evaluate the costs of our opaque predicates, we observe the change in the number of \acp{LUT} and \acp{FF} when applying our two implementation strategies to \ac{FPGA} designs.
To this end we implemented two case studies where we obfuscate Trojans in cryptographic cores with our techniques to harden against detection via reverse engineering.
Designers are afraid of purposely implemented backdoors in designs since the crypto wars in the 1990s and this fear has been rekindled by the Snowden revelations in 2013.
Hence we regard this scenario as a realistic target for obfuscation.

The first case study targets a PRESENT encryption core with input triggered Trojan and the second case study an RSA core, subverted with a kleptographic backdoor.
Although we executed our case studies on \acp{FPGA}, our technique is in no way exclusive to this architecture but can also be used on \acp{ASIC}.

Our expectations are that for the instantiation in existing circuitry, we observe no change in the number of \acp{LUT} and \acp{FF} while for the approach with additional circuitry, the amount of \acp{FF} increases by $n$ and the amount of \acp{LUT} slightly increases because of the feedback logic and routing.

In the following we briefly introduce the designs before presenting the measurement results.

\subsubsection{Trojanized PRESENT}
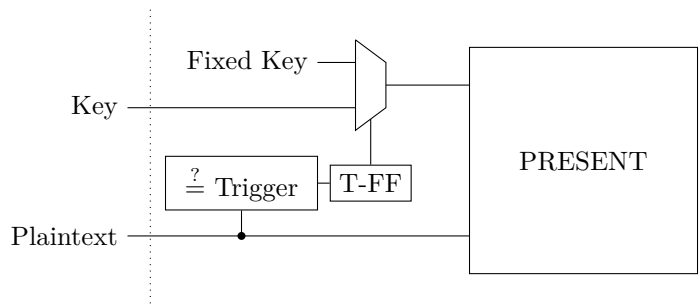
\begin{figure}[!htb]
    \centering

    \begin{tikzpicture}[x=1cm,y=1cm]

    %\draw [help lines] (0,0) grid (4,4);
    \node[rectangle, draw,minimum width=3cm,minimum height=3cm] (present) at (3.5,2) {PRESENT};
    \node[rectangle, draw,minimum width=2cm] (comp) at (-1,1.7) {$\stackrel{?}{=}$ Trigger};
    \node[rectangle, draw] at (0.7,1.7) (TFF) {T-FF};
    \node[anchor=east] at(0,3.3) {Fixed Key};
    \node[anchor=east] at(-2.5,2.7) {Key};
    \node[anchor=east] at(-2.5,1) {Plaintext};

    \draw (0,3.3) -- (0.5,3.3);
    \draw (-2.5,2.7) -- (0.5,2.7);
    \draw (0.5,3.6) -- (0.5,2.4) -- (0.9, 2.7) -- (0.9, 3.3) -- (0.5,3.6);
    \draw (0.9, 3) -- (2,3);

    \draw (-2.5,1) -- (2,1);
    \draw (comp) -- (TFF) -- (0.7,2.55);
    \draw (-1,1) -- (comp);
    \fill (-1,1) circle [radius=1.5pt];

    \draw[dotted] (-2.2,4) -- (-2.2,0);

    \end{tikzpicture}
    \caption{Visualization of the Trojanized PRESENT core.}
    \label{op::fig::eval::present-trojan}
\end{figure}

The first case study features a trojanized PRESENT encryption core as illustrated in \Cref{op::fig::eval::present-trojan}.
The general Trojan idea is similar to the design in \cite{swierczynski:2016:jce}.
When active, the Trojan replaces the user-supplied key with a fixed embedded key.
To an observer, the output still looks encrypted, and multiple infected devices still successfully communicate.
The scenario that multiple devices are infected is likely, if the Trojan is included in the \ac{IP}-core that was used when designing the device family.

Thus, with knowledge of the trigger and the fixed key, an attacker can decrypt and manipulate any traffic.
The Trojan is input-toggled, i.e. once the trigger plaintext is received the Trojan stays active until it is presented a second time with the trigger plaintext.
Note that this Trojan can be used by anyone who has knowledge of the trigger value and the fixed key, so the attacker does not have to be the Trojan designer.
We use opaque predicates to obfuscate the fixed key and the trigger value.
We emphasize, that this protects only the constant key and trigger value, not the Trojan or its trigger mechanism itself, against detection.
For example, Trojan detection techniques such as MERO by Chakraborty~\etal \cite{chakraborty2009mero} still apply.
Via statistical extraction of rarely active paths and automatic test pattern generation (ATPG) the goal of MERO is to generate a minimal test set which stimulates each of these rare paths multiple times.
During dynamic analysis, while feeding the inputs of that test set, triggering a Trojan is highly probable.
Regardless of our opaque predicates, this approach would still detect our Trojan.
Intuitively, an obfuscation technique which transforms rarely active paths to paths with high activity could potentially defend against detection via tools like MERO.
We leave this as an open question for further research.

\paragraph{Obfuscation Impact.}
Since we protected the constant values in the design, we focus on an reverse engineer who tries to extract the trigger value and the fixed key or performs a quick scan for information which hints unexpected constant values in the original HDL code.
For example, in order to find the trigger value, the reverse engineer could scan for comparators with fixed values.
For the fixed key, he could search for an isolated control signal which is spread to all data inputs of a register of the same size as the key.
Our opaque predicates render any kind of attack ineffective which focuses on information introduced by constants in HDL code.
For example, the obfuscation forces the synthesis toolchain to turn the comparator with a fixed value into a dynamic comparator, removing any indication about a constant trigger value.
The adversary could still find a comparator, but to make sure that this is a fixed-input trigger or to extract the trigger value, further reversing is required.

\paragraph{Implementation.}
We implemented the Trojan three times, once unobfuscated, i.e. as depicted in \ref{op::fig::eval::present-trojan}, once using an existing \ac{FSM} and once applying our strategy with additional circuitry, generated via the \texttt{RND} algorithm.
A simple UART core was added to handle I/O.
For the approach with additional circuitry, we generated an opaque predicate of size 3 bits as this was the average size of \ac{FSM} state registers in the design.
We synthesized, placed and routed the design for a Xilinx Artix 7 35T \ac{FPGA}.
The optimization strategy was set to optimize for area and the design was fully flattened.

\subsubsection{Subverted RSA}
The second case study features an RSA core where key generation is subverted with a kleptographic backdoor.
Kleptography has been introduced by Young and Yung around 1996 \cite{young:1996:icc,young:1997:ec}.
In recent years there has been numerous research which presents kleptographic Trojans on cryptosystems \cite{young2007space,gogolewski2008kleptographic,gogolewski2006kleptographic,cryptoeprint:2017:kwant}.
Kleptography describes the study of subliminally and securely stealing information.
The output of a device with a kleptographic Trojan remains conform to specifications when observed with black-box access, but leaks sensitive information exclusively to the Trojan designer.
To this end, it asymmetrically encrypts (parts of) the sensitive information with an embedded public key of the designer, leaking this so called ``kleptogram'' over one or multiple outputs.
In this case study we use our opaque predicates to obfuscate such an embedded key.

We implemented our own kleptographic Trojan which replaces the key generation of the RSA parameters $(n,e,d)$ as shown in \Cref{op::alg::klepto-rsa}.
It uses the Trojan designer's RSA public key $(N_{adv}, E_{adv})$ to generate kleptograms.
The advantage of this approach is that kleptograms are generated without requiring any additional hardware modules.
Only small changes to the datapath and a single \ac{FSM} were enough to subvert the original core.
This is a difference to existing kleptographic attacks on RSA, where additional modules, for example a dedicated hash function, were required.
However, our kleptographic Trojan only works on RSA key generation where $e$ is not fixed but randomly chosen.

\begin{algorithm}[!htb]
    \caption{Subverted RSA KeyGen}
    \label{op::alg::klepto-rsa}
    \begin{algorithmic}[1]
        \Require{$1^\lambda$}
        \Ensure{pk = $(n,e)$, sk = $(d)$}

        \State{Choose $p,q$ as random $\lambda/2$-bit primes}
        \State{$n \gets pq$}
        \State{$e \gets p^{E_{adv}} \mod N_{adv}$}
        \While{gcd$(e, \Phi(n)) \neq 1$}
        \State{$e \gets e + 1$}
        \EndWhile

        \State{$d \gets e^{-1} \mod \Phi(n)$}

        \State{\textbf{return} pk = $(n,e)$, sk = $(d)$}
    \end{algorithmic}
\end{algorithm}

Instead of randomly choosing the public exponent $e$, it is computed as $e = p^{E_{adv}} \mod N_{adv}$ and then incremented until it is invertible modulo $\Phi(n)$.
Subsequently, the Trojan designer has to compute $p' = (e-i)^{D_{adv}} \mod N_{adv}$ for a small number of i, until $n$ is divisible by $p'$.
Then he can factorize $n$ as $p = p'$, $q = n/p$, break RSA, and thus decrypt or even manipulate all ongoing communication.

\paragraph{Obfuscation Impact.}
Note that in contrast to the PRESENT case study, our kleptographic Trojan is always active.
Techniques for Trojan detection often aim at finding triggers or circuitry which is inactive most of the time (c.f. \cite{chakraborty2009mero, waksman:2013:ccs, zhang:2013:dac}) and are therefore not capable of detecting kleptographic Trojans.

We try to protect against a reverse engineer who tries to detect a kleptographic Trojan by using the knowledge that the design has to include the Trojan designer's public key.
As the key pair $(N_{adv}, E_{adv})$ has to be included in the design, he could try to locate constant signals which form a possible input to the exponentiation engine.
We use an opaque predicate to obfuscate $N_{adv}$ and $E_{adv}$ in order to harden against this detection approach, as all inputs to the exponentiation engine now appear to be dynamic without deep inspection.

\paragraph{Implementation.}
We selected an exemplary 256-bit RSA-core, capable of encryption, decryption, and key generation.
Again, we implemented the design three times: unobfuscated and once with each of our two techniques, using the \texttt{RND} algorithm for our approach with additional circuitry.
We use the same UART core as in the previous case study to handle I/O.
When applying our first implementation technique we used a single \ac{FSM} to generate both values $N_{adv}$ and $E_{adv}$.
For the approach with additional circuitry, we generated an opaque predicate of size 5 bits as this was the average size of \ac{FSM} state registers in the design.
We synthesized, placed and routed the design for a Xilinx Artix 7 35T \ac{FPGA}, to get comparable results to the first case study.
The optimization strategy was set to optimize for area and the design was fully flattened.

\subsubsection{Implementation Results}
\label{op::sec::eval::results}

\begin{table}[!htb]
    \centering
    % not fully flattened
    \begin{tabular}{cl|rl|rl}
    	      \multicolumn{2}{c|}{Design}       & \multicolumn{2}{c|}{LUTs} & \multicolumn{2}{c}{FFs} \\ \hline
    	\multirow{3}{*}{PRESENT} & Unobfuscated &   304 &                   &  347 &                  \\
    	                         & Strategy 1   &   307 & +0.99\%           &  347 & +0\%             \\
    	                         & Strategy 2   &   304 & +0\%              &  350 & +0.86\%          \\ \hline
    	\multirow{3}{*}{RSA}     & Unobfuscated & 10570 &                   & 5316 &                  \\
    	                         & Strategy 1   & 10811 & +2.28\%           & 5314 & $-0.04\%$        \\
    	                         & Strategy 2   & 10692 & +1.15\%           & 5323 & +0.13\%
    \end{tabular}

    \caption{Area comparison of our three case studies. Percentual changes refer to the unobfuscated circuit.}
    \label{op::tab::eval}
\end{table}

\Cref{op::tab::eval} shows the results of our evaluation in terms of \ac{LUT} and \ac{FF} count.
It contains the absolute numbers for each variant of the two case studies as well as the relative increase compared to the unobfuscated design.
\textit{Strategy 1} refers to the implementation which uses an existing \ac{FSM}, and \textit{Strategy 2} to the one where new circuitry for the opaque predicate is added.
We confirmed manually that the additional circuitry remained untouched by the synthesizing toolchain, which is another indicator that they are difficult to detect via static analysis, since compiler optimization is based on the results of a static analysis.
For \textit{Strategy 2}, we connected the reset signals of our opaque predicates to reset signals of \acp{FSM} which were reset far earlier in both case studies.
Therefore, our obfuscations have no impact on latency as the opaque predicates had time to stabilize before the processing period.
Although \textit{Strategy 1} is expected to have no impact on area, as we solely use existing circuitry, the number of \acp{LUT} increases by 0.99\% for the PRESENT case study and by 2.28\% for the RSA core.
We expect this increase to result from layout optimization of the synthesis toolchain to fit the \ac{FPGA} architecture.
Regarding the amount of \acp{FF}, our expectations are met, as no additional \acp{FF} were instantiated.
In the RSA case study, the optimizer managed to take advantage of the new connections and reduced the number of \acp{FF} by 0.04\% compared to the unobfuscated design.

For \textit{Strategy 2} we expected a slight increase in both, \ac{FF} and \ac{LUT} count.
For the PRESENT case study there is an increase of exactly 3 \acp{FF}, i.e. the \acp{FF} which were added for the opaque predicate.
However, in the RSA case study, 7 new \acp{FF} were added, although the opaque predicate only uses 5.
We assume that the additional a \acp{FF} are internal buffers, also inferred by the optimization tools.
Surprisingly the amount of \acp{LUT} did not increase for the PRESENT core.
Apparently, the additional logic can be embedded into underutilized \acp{LUT}.
In the RSA case study the number of \acp{LUT} increases by 1.15\% which reflects our expectation, although the increase exceeds the required gates for the $f$ function.

Our evaluation shows that the theoretical expectations are not exactly met in reality.
Although the impact on the amount of \acp{FF} is as expected, the number of \acp{LUT} increases more than anticipated.
Optimizations applied by the synthesis toolchain heavily influence the numbers.
Note that the toolchain itself was not able to deduce that our construction actually generates a constant value and did not replace it.
While we did not achieve a zero-overhead obfuscation using our implementation strategies, the results still attest very low overhead.

%!TEX root = ../main.tex

\section{Application to Watermarking}
\label{op::sec::outlook}
In the previous section we presented two case studies which focused on malicious application of obfuscation techniques to hide a Trojan.
In this section we present the theory of a defensive application of our opaque predicates, where a designer tries to protect intellectual property.

We apply our obfuscation to the watermarking scheme of Schmid~\etal \cite{schmid:2008:fpt}, effectively rendering previous removal attacks ineffective.
The watermarks of Schmid~\etal are used to protect \ac{IP}-cores on \ac{LUT}-based \acp{FPGA} and can be verified from the bitstream file that is configuring the \ac{FPGA}.
The configuration of a \ac{LUT} in the bitstream file defines the output values for each input combination.
Subsequently, for an $n$-to-$1$ bit \ac{LUT} $2^n$ bits are required for configuration.
The idea of Schmid~\etal was to fix some inputs of a \ac{LUT} to 0, i.e. GND.
Therefore, all input combination where those inputs would be 1 cannot occur during operation.
The bits in the \ac{LUT} configuration which are intended to configure the corresponding output bits are then used to encode a watermark.
As watermarks may not fit in one \ac{LUT} configuration, it can be split over multiple \acp{LUT}.

An example is illustrated in \Cref{op::fig::watermark} which shows the truth table and resulting configuration for a 4-to-1 \ac{LUT}.
The inputs $I_3$ and $I_2$ are fixed to GND, so only the input combinations for the four bits labeled \texttt{C} can actually occur during operation.
We marked all other input combinations which cannot occur gray.
For example, the input combination $(I_3, \ldots I_0) = 1000$ cannot occur as $I_3$ is constantly 0.
The 12 configuration bits that belong to these inputs, labeled \texttt{W}, can be used to encode a watermark.
\begin{figure}[!htb]
    \centering
    \begin{tabular}{r|l}
	GND $\rightarrow I_3$ & \texttt{0 0 0 0 \color{gray}{0 0 0 0 1 1 1 1 1 1 1 1}} \\
	GND $\rightarrow I_2$ & \texttt{0 0 0 0 \color{gray}{1 1 1 1 0 0 0 0 1 1 1 1}} \\
	                $I_1$ & \texttt{0 0 1 1 \color{gray}{0 0 1 1 0 0 1 1 0 0 1 1}} \\
	                $I_0$ & \texttt{0 1 0 1 \color{gray}{0 1 0 1 0 1 0 1 0 1 0 1}} \\ \hline
	        output/config & \texttt{C C C C W W W W W W W W W W W W}
    \end{tabular}
    \caption{Example of a 4-to-1 bit \ac{LUT} configuration where the \texttt{C} bits are actual configuration and the \texttt{W} bits can be used for watermarking.}
    \label{op::fig::watermark}
\end{figure}

The bitstream file used to configure the \ac{FPGA} contains all \ac{LUT} configurations and thus the watermark's existence can be verified via exhaustive search.
Removal of the watermarks on bitstream level is considered infeasible, since no information on interconnection is plainly available.
Removal attacks on the netlist level are regarded as far more likely.
Therefore, Schmid~\etal include an attacker, who has access to the netlist of the design, into their attacker model.

In 2017 Wallat~\etal presented a practical removal attack on that scheme \cite{wallat:2017:ivsw}.
With access to the netlist and the information that unused \ac{LUT} inputs are connected to GND, tracing all GND connections to \acp{LUT} instantly reveals the watermarked instances.
Subsequently removal of the watermark is a straightforward task.

To counter removal attacks, Schmid~\etal already proposed the insertion of bogus cells that generate constant signals.
Wallat~\etal directly propose the use of opaque predicates to render their attack useless but also highlight the lack of hardware opaque predicates.

Using our instantiations one can efficiently harden the watermarks of Schmid~\etal against netlist level removal attacks.
The idea is to connect the fixed \ac{LUT} inputs to an obfuscated constant that outputs a zero during the processing period instead of connecting directly to GND.
Since this connection does not affect \ac{LUT} configuration the watermark detection algorithm remains unaltered, but attacks on netlist level are hardened tremendously, since state machine signals that reach \acp{LUT} are very common, i.e. our integration is stealthy.
Furthermore, by using multiple opaque predicates an attacker might not be able to fully remove the watermark.
This can be especially helpful if the watermark is embedded with redundancy.

%!TEX root = ../main.tex

\section{Conclusion}
\label{sec::conclusion}
In this work we presented a novel technique to instantiate opaque predicates in hardware on HDL level.
We described how existing \acp{FSM} or counters can be used to implement obfuscated constants without adding any additional gates.
For situations where no existing circuitry can be used, we presented and compared three algorithms to automatically generate opaque predicates disguised in new \ac{FSM}-like structures.
We substantiated the stealthiness of our approach by taking the viewpoint of an attacker and performed two case studies on an \ac{FPGA} to analyze the impact of our techniques on the overall design size and latency.
The results showed that it is possible to implement our opaque predicates without additional \acp{FF} and only small overhead in \acp{LUT} of approximately 1\%-2.2\% while not impacting latency.
Finally we showed how our technique is able to mitigate the open problem of netlist-level removal attacks for the \ac{LUT}-based IP protection by Schmid~\etal \cite{schmid:2008:fpt}.
To the best of our knowledge, by instantiating opaque predicates using existing \acp{FSM} or counters, our approach is more stealthy than previous work and introduces less overhead.

\section*{Acknowledgments}
We thank the anonymous reviewers for their valuable and helpful feedback.
Furthermore, we would like to thank Pascal Sasdrich for his editorial support.
This work was supported in part by grant ERC 695022 and NSF CNS-1563829.

\bibliographystyle{IEEEtran}
{
  \bibliography{localbib,bibliography}
}

\end{document}